\documentclass[english]{elsarticle}
\usepackage{ae,aecompl}
\usepackage[T1]{fontenc}
\usepackage[utf8]{luainputenc}
\usepackage{geometry}
\usepackage{babel}
\usepackage{array}
\usepackage{float}
\usepackage{rotfloat}
\usepackage{multirow}
\usepackage{amsthm}
\usepackage{amsmath}
\usepackage{amssymb}
\usepackage{graphicx}
\usepackage[unicode=true]
 {hyperref}

\makeatletter

\providecommand{\tabularnewline}{\\}

\renewcommand{\fnum@figure}{Figure~\thefigure}

\renewcommand{\fnum@table}{Table~\thetable}

\makeatletter
\def\ps@pprintTitle{%
  \let\@oddhead\@empty
  \let\@evenhead\@empty
  \def\@oddfoot{}
  \let\@evenfoot\@oddfoot
}
\makeatother

\hypersetup{%
    pdfborder = {0 0 0}
}

\makeatother

\begin{document}
\begin{frontmatter}

\title{{\Large{}Bank Networks from Text:}\\
{\Large{}Interrelations, Centrality and Determinants}\tnoteref{mytitlenote}}

\tnotetext[mytitlenote]{This paper is accompanied by supplementary interactive interfaces:
\href{http://risklab.fi/demo/textnet/}{http://risklab.fi/demo/textnet/}
(for a further discussion of the VisRisk platform see Sarlin \cite{Sarlin2013SWIFT}).
The authors want to thank Tuomas Peltonen and three anonymous reviewers
for insightful comments and discussions. The paper has also gained
from presentations of previous versions of it at the following conferences:
the 11th International Symposium on Intelligent Data Analysis (IDA'12)
on 25--27 October 2012 in Helsinki, Finland, the 19th IEEE Conference
on Computational Intelligence for Financial Engineering \& Economics
(CIFEr'14) on 27--28 March 2014 in London, UK, and Arcada Seminar
on Current Topics in Business, IT and Analytics (BITA'14) on 13 October
2014 in Helsinki, Finland. The first author gratefully acknowledges
the Graduate School at Åbo Akademi University and the second author
the GRI in Financial Services, Louis Bachelier Institute, and the
Osk. Huttunen Foundation for financial support. Corresponding author:
Samuel Rönnqvist, Turku Centre for Computer Science, Department of
Information Technologies, Åbo Akademi University, Turku, Finland.
E-mail: sronnqvi@abo.fi.}

\author[S1]{Samuel Rönnqvist}

\author[P1,P3]{and Peter Sarlin}

\address{\smallskip{}
}

\address[S1]{Turku Centre for Computer Science -- TUCS, \\
Department of Information Technologies at Åbo Akademi University,
Turku, Finland\\
\medskip{}
}

\address[P1]{RiskLab Finland at Arcada University of Applied Sciences, Helsinki,
Finland\\
\medskip{}
}

\address[P3]{Department of Economics at Hanken School of Economics, Helsinki,
Finland}
\begin{abstract}
In the wake of the still ongoing global financial crisis, bank interdependencies
have come into focus in trying to assess linkages among banks and
systemic risk. To date, such analysis has largely been based on numerical
data. By contrast, this study attempts to gain further insight into
bank interconnections by tapping into financial discourse. We present
a text-to-network process, which has its basis in co-occurrences of
bank names and can be analyzed quantitatively and visualized. To quantify
bank importance, we propose an information centrality measure to rank
and assess trends of bank centrality in discussion. For qualitative
assessment of bank networks, we put forward a visual, interactive
interface for better illustrating network structures. We illustrate
the text-based approach on European Large and Complex Banking Groups
(LCBGs) during the ongoing financial crisis by quantifying bank interrelations
and centrality from discussion in 3M news articles, spanning 2007Q1
to 2014Q3.\end{abstract}
\end{frontmatter}

\newpage{}

\section{Introduction}

The global financial crisis has brought several banks, not to say
entire banking sectors, to the verge of collapse. This has not only
resulted in losses for investors, but also costs for the real economy
and welfare at large. Considering the costs of banking crises, the
recent focus of research on financial instabilities is well-motivated.
First, real costs of systemic banking crises have been estimated to
average at around 20--25\% of GDP (e.g., \cite{dell2008real,laeven2010resolution}).
Second, data from the European Commission illustrate that government
support for stabilizing banks in the European Union (EU) peaked at
the end of 2009. The support amounted to  Euro 1.5 trl, which is more than
13\% of EU GDP. The still ongoing financial crisis has stimulated
a particular interest in systemic risk measurement through linkages,
interrelations, and interdependencies among banks. This paper advances
the literature by providing a novel measure of bank linkages from
text and bank importance through information centrality.

Most common sources for describing bank interdependencies and networks
are based upon numerical data like interbank asset and liability exposures
or payment flows, and co-movements in market data (e.g., equity prices,
CDS spreads, and bond spreads) (see \cite{cerutti2012systemic}).
While these direct and indirect linkages complement each other, they
exhibit a range of limitations. Even though in an ideal world bank
networks ought to be assessed through direct, real linkages, interbank
data between banks' balance sheets are mostly not publicly disclosed.
In many cases, even regulators have access to only partial information,
such as lack of data on pan-European bank linkages despite high financial
integration. In this vein, a commonly used source of data descends
from interbank payment systems (see \cite{Soramakietal2007}), but
is again only accessible for a limited set of regulators. It is also
worth to note that real exposures, as they are measured for individual
markets, are oftentimes highly biased towards the business model of
a bank, such as investment or depository functions. Market price data,
while being widely available and capturing other contagion channels
than those in direct linkages between banks \cite{acharya2012measuring},
assume that asset prices correctly reflect all publicly available
information on bank risks, exposures and interconnections. Yet, it
has repeatedly been shown that securities markets are not always efficient
in reflecting information about stocks (e.g. \cite{malkiel2003efficient}).
Further, co-movement-based approaches, such as that by Hautsch et
al \cite{hautsch2013financial}, require large amounts of data, often
invoking reliance on historical experience, which may not represent
the interrelations of today. Also, market prices are most often contemporaneous,
rather than leading indicators, particularly when assessing tail risk.
It is neither an entirely straightforward task to separate the factors
driving market prices in order to observe bilateral interdependence
\cite{borio2009towards}.

Big data has emerged as a central theme in analytics during the past
years. Research questions of big data analytics arise not only from
massive volumes of data, or speeds at which data are constantly generated,
but also from the widely varying forms, particularly unstructured
textual data, that in themselves pose challenges in how to effectively
and efficiently extract meaningful information \cite{dhar2013data}.
This paper treats the text mining aspect, as it proposes an approach
to relationship assessment among banks by analyzing how they are mentioned
together in financial discourse, such as news, official reports, discussion
forums, etc. The idea of analyzing relations in text is in itself
simple, but widely applicable. It has been explored in various areas;
for instance, Özgür et al. \cite{ozgur2008co} study co-occurrences
of person names in news, and Wren et al. \cite{wren2004knowledge}
extract biologically relevant relations from research articles. These
approaches can be used to construct social or biological networks,
using text as the intermediate medium of information. Our contribution
lies in proposing this text-based approach to the study of bank interrelations,
with emphasis on analysis of the resulting bank network models and
ultimately quantifying a bank's importance or centrality.

Our approach may be compared to the above discussed, more established
ways of quantifying bank interdependence, such as interbank lending
and co-movement in market data. While not measuring direct interdependence,
it has the advantage over interbank exposures by relying upon widely
available data, and over co-movements in market data by being a more
direct measure of an interrelation. By contrast, our approach serves
to shed light on banks' relationships in the view of public discussion,
or of information overall, depending on the scope of textual data.
It may serve as a way of tapping into the wisdom of the crowd, while
offering a perspective different from previous methods, especially
considering the presence of rich, embedded contextual detail. Rather
than an ending point, this sets a starting point from which further
study may focus more extensively on the context of occurrences and
more sophisticated semantic analysis. This allows to better understand
factors driving interrelations, and overall centrality.

In this paper, we assess European Large and Complex Banking Groups
(LCBGs) using the text-based approach for quantifying bank interrelations
from discussion in the news. A co-occurrence network is derived from
3 million articles, published during 2007Q1 to 2014Q3 in the Reuters
online news archive. Beyond only quantifying bank interrelations,
we also provide means for quantitative and qualitative assessment
of networks. To support quantification of bank importance, we propose
an information centrality measure to rank and assess trends of bank
centrality in discussion, which relates to the information channel
in the analysis of interconnected, and potentially systemic, financial
risk. In contrast to common shortest-path-based centrality measures,
information centrality captures effects that might propagate aimlessly
by accounting for parallel paths. Thus, rather than direct financial
exposures, we provide a representation of the channel for potential
informational contagion, as well as other common factors leading to
co-occurrence in discussion, such as overlapping portfolios and exposure
to common exogenous shocks. To support a qualitative assessment of
the bank networks, we put forward a visual, interactive interface
for better illustrating network structures. This concerns not only
an interface to network models, but also an interactive plot to better
communicate quantitative network measures.%
\footnote{The interactive interfaces are provided as web-based implementations:
\href{http://risklab.fi/demo/textnet/}{http://risklab.fi/demo/textnet/}.
For a further discussion of the VisRisk platform see Sarlin \cite{Sarlin2013SWIFT}.%
}

The co-occurrence network illustrates relative prominence of individual
banks, and segments of more closely related banks. The systemic view
acknowledges that the centrality of a bank in the network is a sign
of importance, and not necessarily its size (cf. too central to fail
by Battistone et al. \cite{Battistonetal2012}). The dynamics of the
network, both local and global, reflect real-world events over time.
The network can also be utilized as an exploratory tool that provides
an overview of a large set of data, while the underlying text can
be retrieved for more qualitative analysis of relations. 

To better understand what drives information centrality, and how it
ought to be interpreted, we explore determinants of the centrality
measure. We investigate a large number of bank-specific risk drivers,
as well as country-specific macro-financial and banking sector variables,
as well as control for variables measuring bank size. Further, we
also assess the extent to which information centrality explains banks'
risk to go bad, and compare it to more standard measures of size.
Even though bank size is a key factor explaining information centrality,
we show that centrality is not a direct measure of vulnerability.
This implies that the centrality measure is not biased by the nature
of business activities or models, which potentially impacts bank vulnerability
(e.g., asset size or interbank-lending centrality). Rather than a
narrow, direct measure of interconnectedness, we are capturing systemic
importance of a bank more broadly, in terms of connectivity expressed
in financial discourse. Yet, while the rich nature of textual data
provides possibilities to more specifically query and define interrelationships
and other potentially interesting details on banks, interpreting the
semantics by computational methods is often challenging. To this end,
we also discuss different ways of analyzing text-based networks, laying
forward some ideas on future directions in their study.

The following section explains the data and methods we use to construct
and analyze bank networks from text, whereas Section 3 discusses the
results of the experiments on textual data, including both qualitative
and quantitative analysis. Before a concluding discussion on text-based
networks, Section 4 assesses determinants of information centrality.

\section{Bank networks from text: Data and methods}

This section provides a discussion of the text-to-network process,
both generally and from the viewpoint of the study in this paper.
First, we detail the particular text data and choice of banks to be
studied. Having established this, we turn to the process of text analysis
and construction of bank co-occurrence networks. This is followed
by discussion on the analysis of such networks, including both quantitative
and qualitative analysis.

\subsection{Data and target banks}

Through digitized economic, social and academic activities, we are
having access to ever increasing amounts of textual data. While vast
amounts of textual data are readily available, there is nothing that
assures increases in precision and quality of data. Analytics of big
data is increasingly a search for needles in haystacks, where choices
in data source, collection methods as well as pre-processing setups
all need to be carefully directed in order to pick up desired signals.
Likewise, when tapping into financial discourse, one needs to clearly
narrow the context of collected data and targeted entities of interest,
beyond the choice of data source.

\begin{figure}
\begin{centering}
\includegraphics[width=0.9\textwidth]{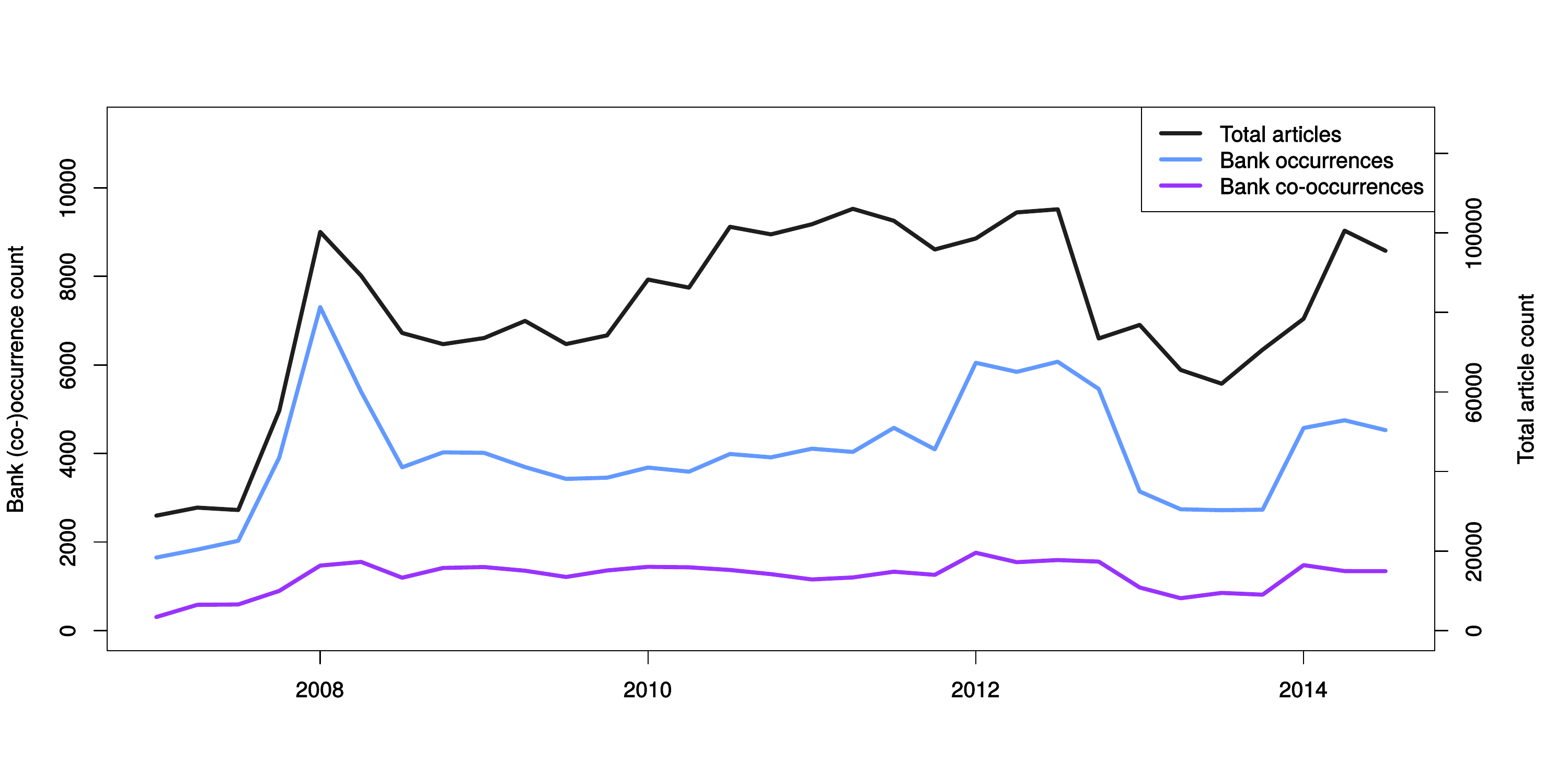}
\par\end{centering}

\protect\caption{Volumes of all news articles and bank name occurrences over time.\label{fig:Volumes-of-all}}
\end{figure}

The text data we use in this paper is newly collected from Reuters
online news archive. News text presents a rather formal type of discourse,
which eases interpretation of extracted relations, as opposed to more
free-form, user-generated online discussion as explored in earlier
work by Rönnqvist and Sarlin \cite{RonnqvistSarlin2014}. We focus
on major consumer banks within Europe, classified by the European
Central Bank \cite{ECB13fsr} as Large and Complex Banking Groups
(LCBGs), of which 15 are also classified as Globally Systemically
Important Banks (G-SIBs) by the Financial Stability Board \cite{FSB2013update}.
See Table A.1 in the Appendix for a list of LCBGs and G-SIBs and the
naming convention used in this paper. The period of study is 2007Q1--2014Q3,
for which the news archive contains 6.7M articles. We base our analysis
on a 45\% random sample of articles comprising of 3.0M articles (1.5B
words).

The text analysis is based on detecting mentions of bank names in
the articles. We look at a set of 27 banks: 5 British, 5 French, 4
German, 4 Spanish, 3 Dutch, 2 Italian, 2 Swiss, 1 Swedish and 1 Danish
bank. In order to mitigate a geographical sampling bias, we use the
U.S. edition of the Reuters news archive, as no single European edition
is available, but rather national editions for only the largest countries.

The chart in Figure \ref{fig:Volumes-of-all} provides an overview
of the trends in total news article volume, as well as the volume
of bank name occurrences. Out of all articles, 5.4\% mention any of
the targeted banks, on average. The volume is relatively low in the
beginning of 2007, i.e., the start of the archive. Mentions of banks
reaches a peak in early 2008, after which it fluctuates between 60k
and 110k articles per quarter.

\subsection{From text to bank networks}

\begin{figure}
\begin{centering}
\includegraphics[width=5.5in]{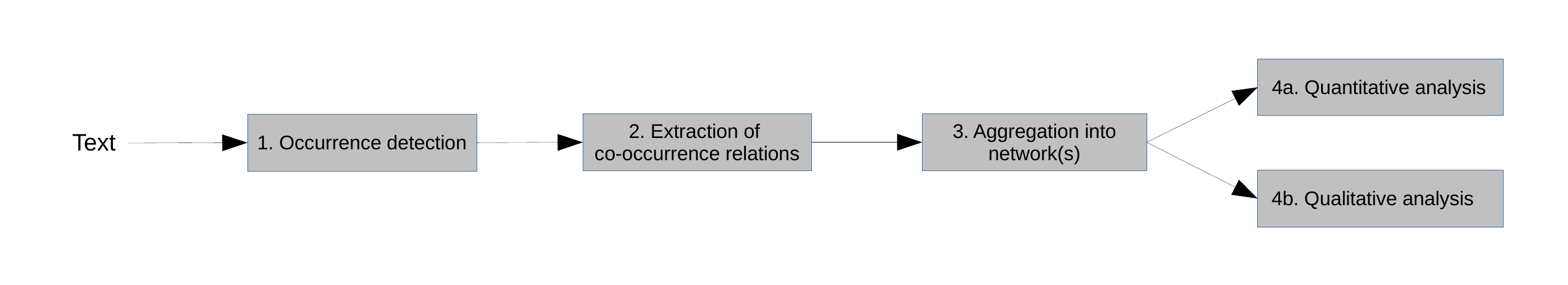}
\par\end{centering}

\protect\caption{Text-to-network process: (1) Occurrences of bank names are detected
in source text, (2) pair-wise co-occurrence relations are extracted
between occurrences within a context, and (3) relations aggregated
over a time interval form a co-occurrence network. A resulting network
can be analyzed with (4a) quantitative measures capturing some interesting
features, and (4b) qualitative analysis through visual exploration
of the network, its neighborhoods, and connectivity of individual
nodes.\label{fig:From-text-to-network}}
\end{figure}

With plain text as a starting point, and relationship assessment as
an objective, we analyze co-mentions in financial discourse. Extracting
occurrences and co-occurrences from text is the initial step. The
relationships are constituents of co-occurrence networks, whose properties
can be assessed through both quantitative and visual analysis. Figure
\ref{fig:From-text-to-network} provides an overview of the process
of transforming text into network models that lend themselves to analysis.

To construct the network we scan the text for occurrences of bank
names to detect and register mentions of those banks. Scanning is
performed using patterns manually designed and tested to match with
as high accuracy as possible. Generally, the use of manually designed
patterns for information extraction in text tend to have high precision
but lower recall, but we expect that the reasonably standardized form
of discourse we use should mitigate a loss in recall. The pattern
for each bank is specified as a set of regular expressions targeting
common naming variants such as full name, abbreviations, synonymous
names, names of subsidiaries, historical names and spelling variations.
The regular expressions are developed and tested iteratively on data
to optimize accuracy, going from broader patterns toward higher precision
with retained recall.

Co-occurrence analysis is computationally very efficient and versatile
in terms of language, compared to more technically sophisticated relation
extraction techniques (e.g., based on dependency parsing \cite{bunescu2005shortest}),
while it offers worse precision of relations. Using co-occurrence-based
relation extraction lets us process billions of words on standard
architecture serially in the order of hours, and we assume the substantial
data volume to partially compensate for the noisier relation extraction.
The framework is language independent and works equally well on English
language news as, for instance, on Finnish online discussion \cite{RonnqvistSarlin2014}.

A co-occurrence relation is formed by two bank names occurring in
the same context. In the present case, we define the scope of the
context as a 400-character sliding window in the text, whereas a wider
scope would require less data but increase noise as any co-occurrence
is less likely to represent a meaningful relation. In the process,
a context is checked for co-occurrence candidates as follows. A context
is scanned for substrings matching the defined regular expressions,
and a bank occurrence is registered by associating the matching pattern
with its corresponding bank. Multiple occurrences of a single bank
are counted only once per context, ignoring presumed meaningless repetitions,
but an occurrence may participate in multiple relations. A context
containing two or more banks yields one or more pair-wise co-occurrence
relations. Thus, derived from the set of matches \emph{$M\subset\mathbb{N}$
}(indexed by bank) in context\emph{ c, }we define the set of co-occurrence
relations \emph{R }as:

\[
R_{c}=\left\{ r\mid r\in M_{c}\times M_{c}\wedge r_{1}<r_{2}\right\} 
\]

\medskip{}

However, we disqualify contexts with more than 5 banks, as they are
likely to be listings that would result in marginally meaningful relations.
These design decisions should be adjusted and tested for each new
data source, to obtain less noisy results.

We aggregate co-occurrences over time to form links that are weighted
by the absolute co-occurrence count during a period (e.g., a quarter).
These links form a dynamic network, a series of cross-sectional networks,
which allows the extracted relations to be studied using methods for
analysis of complex networks. In the network, banks form nodes (or
vertices), and aggregated co-occurrence relations form their links
(or edges). To extract meaningful quantitative measures of co-occurrence
networks, measures designed for weighted networks need to be used.
Nevertheless, most conventional network analysis methods are designed
for binary (unweighted) networks only \cite{opsahl2010node}, which
calls for some form of transformation of the network if these measures
are to be used, such as by filtering out very weak connections. While
unfiltered networks are more sensitive to noise when using binary
measures, low-frequency co-occurrences may be of particular interest,
as they are more likely to represent novel information. In order not
to lose detail, it is highly motivated to use weighted networks and
measures that account for link weights. Larger sample size or longer
aggregation intervals increase the co-occurrence count, i.e., the
weights of the cross-sectional networks, and will affect many network
measures (including information centrality discussed later); as the
networks are directly comparable among cross sections this is however
not an issue.

Although quantitative analysis of networks provides means to better
understand overall properties of networks, they as any aggregate measure
most often lack in detail. Hence, network visualization supports not
only detailed analysis of network structure and constituents, but
also further details as demanded. In the following subsection, we
further discuss both quantitative measurement of network properties
and visualization as a support in their analysis.

\subsection{Network analysis}

Network models are commonly rather complex and rich in information.
They can be analyzed in many different ways to gain insight into the
nature of the underlying phenomenon, the bank connectivity landscape
in our case. We first discuss analysis of the networks at a global,
descriptive level, to describe properties of the co-occurrence networks
through common network measures. Later, we concentrate on the concept
of centrality and a few ways of quantifying it in our type of network,
with the study of systemic risk in mind. Finally, we discuss network
visualization as a means for interactive exploration.

\subsubsection{Global properties}

A commonly cited property of real-world networks is that the average
distance between nodes is very small relative to the size of the network,
lending them the name ''small-world\emph{''} networks \cite{watts1998collective}.
Short distances have a functional justification in most types of network,
as it increases efficiency of communication, while there also is a
general tendency towards short average distances among non-regular
networks. These networks have varying \emph{degree}, i.e., number
of links per node, the distribution of which is a typical way of profiling
empirical networks. Networks that have evolved through natural, self-organizing
processes, such as communications, social, biological and financial
networks, tend to exhibit degree distributions that follow a power
law. These so-called \emph{scale-free} networks evolve through processes
of preferential attachment, where the likelihood of a node receiving
a new link is proportional to its current degree \cite{barabasi1999emergence}. 

Jackson \& Rogers \cite{jackson2007meeting} distinguish two archetypes
of natural networks, described by power-law degree distributions and
exponential degree distributions, respectively. They argue that, in
fact, empirical networks generally exhibit hybrid distributions, between
power-law and exponential, as they are formed through mixed processes
of preferential attachment and attachment with uniform probability.
The latter process still generates highly heterogeneous exponential
distributions, as established nodes have greater chance over time
at growing well embedded into the network. By either process, some
nodes are bound to be more influential than others, and mapping the
levels of influence in the system is our main interest. To profile
the co-occurrence networks, the average shortest paths and degree
distributions can indicate how small-world and scale-free they are.
In the latter case, as we are interested in accounting for the link
weighting, we study the distribution of strength, i.e., weighted degree
calculated as the sum of weights per node (as \cite{Barrat16032004}
propose).

Other typical ways of characterizing structure focus on network density
and modularity. For instance, a clustering coefficient can measure
the probability that triplets of connected nodes in binary networks
form triangles, providing a measure of density that can be conditioned
on degree, etc. Networks may consist of several modules or communities,
i.e., subnetworks more densely connected to eachother than to other
parts. Although such characteristics can be studied by quantitative
means, it is not of particular interest for the current news-based
bank networks. However, we will briefly consider these qualities based
on visual analysis in Subsection 3.2.

\subsubsection{Centrality}

Following the initial profiling of the whole network, we turn the
focus toward the concept of node centrality. A central node holds
a generally influential position in a network; a centrally located
bank is likely to be systemically important, as it stands to affect
a large part of the network directly or indirectly in case of a shock
(negative or positive). There is, however, a range of ways to quantify
centrality, the most common measures being degree centrality (i.e.,
fraction of nodes directly linked) and the shortest-path-based closeness
centrality and betweenness centrality. We adapt degree centrality
to our weighted networks, by using strength as a direct measure of
centrality. Closeness and betweenness centrality can also incorporate
link weight into the calculation of shortest path, by means of the
Dijkstra's shortest-path algorithm \cite{dijkstra1959note} that interprets
weights as distances between nodes. Since co-occurrence networks represent
tighter connections (i.e., more co-occurrences) by higher weights,
it is necessary to invert the weights before calculation, as proposed
by \cite{newman2001scientific}.

Borgatti \cite{borgatti2005centrality} points out that a common mistake
in the study of network centrality is to neglect to consider how flow
in the system is best modeled. The common shortest-path-based centrality
measures make implicit assumptions that whatever is passing from a
node to the surrounding network does so along optimal paths, such
as in routing networks of goods and targeted communication. Arguably,
a more realistic intuition for influence of a node, in cases where
effects might propagate aimlessly, such as any type of contagion,
is one that accounts for parallel paths that may exist. 

Along these lines, we study a closeness centrality measure that models
the flow of information in such a manner, called \emph{information
centrality} \cite{Stephenson19891} (also known as current flow closeness
centrality \cite{brandes2005centrality}). Information centrality,
which seeks to quantify the information that can pass from a node
to the network over links whose strength determine level of loss in
transmission, is defined as

\begin{equation}
I(i)=\frac{n}{nC_{ii}+\sum_{j=1}^{n}C_{jj}-2\sum_{j=1}^{n}C_{ij}}
\end{equation}
\smallskip{}

where \emph{$n$ }is the number of nodes and the weighted pseudo-adjacency
matrix is defined as

\[
C=B^{-1},\; B_{ij}=\begin{cases}
1+S(i), & \mbox{if}\; i=j\\
1-w_{ij}, & \mbox{otherwise}
\end{cases}
\]

\smallskip{}
where \emph{$w$ }is link weight (0 for unlinked nodes) and \emph{$S(i)$
}is strength of node \emph{$i$. }This allows us to measure the centrality
or influence of bank $i$ in public discourse, which relates to a
very general-purpose measure of connectedness in discussion. When
relating to systemic risk, we aim at capturing the information channel
when analyzing interconnected financial risk. Thus, rather than direct
financial exposures, we provide a representation of the channel for
potential informational contagion, as well as other common factors
leading to co-occurrence in discussion, such as overlapping portfolios
and exposure to common exogenous shocks.

Centrality as a measure of a node's relative importance is interesting,
yet changes in centrality adds another dimension. We study networks
of quarterly cross sections of the data, in order to calculate and
compare centralities over time.

When the data is split by shorter intervals less frequent parts will
inevitably become disconnected from the main network component. Information
centrality is quite sensitive to the resulting fluctuations in component
size, while the more central nodes start to correlate strongly. We
propose a method to stabilize the centrality measurement by applying
Laplace smoothing to the link weights before calculation of information
centrality. The weight of each existing link is increased by a small
constant (e.g. 1.0), while links are added between all other nodes
and weighted by the same constant. Formally, $w'_{ij}=w_{ij}+\alpha$,
where $w_{ij}=0$ if \emph{i} and \emph{j} are not connected. The
reasoning is that operating on a limited sample of links, we want
to discount some probability for unobserved links (between known nodes),
to lessen the influence that the difference between non-occurring
(unobserved) links and single-occurrence links has on centrality.
This type of additive smoothing has similarly been applied in language
modeling \cite{chen1999empirical}, but is generally applicable to
smoothing of categorical data.

The choice of the smoothing parameter $\alpha$ is dependent on the
study objective: modest levels (e.g., 0.1) retain more information
on global changes in centrality, whereas higher levels (e.g., 1.0
or more) accentuate relative differences among nodes. Subsection 3.1
discusses the effects of different levels of smoothing, based on visual
assessment as well as measures of variance. On the one hand, the effect
of smoothing can be quantified through the average variance in information
centrality per node over time periods (\emph{T}):

\begin{equation}
V=\frac{1}{n}\sum_{i=1}^{n}\left(\frac{1}{|T|}\sum_{t\in T}\left(I'_{t}(i)-\mu_{i}\right)^{2}\right)
\end{equation}

\medskip{}
where \emph{$\mu_{i}$ }is mean node centrality over time and \emph{I'
}is smoothed and min-max normalized \emph{I} over all \emph{t} and
\emph{i}. This variance should decrease with increased smoothing.
On the other hand, the relative spread of nodes that is expected to
increase with higher levels of smoothing can be similarly measured
based on variance among nodes in a cross section, rather than among
cross sections for a node. The average is then formulated as:

\begin{equation}
V'=\frac{1}{|T|}\sum_{t\in T}\left(\frac{1}{n}\sum_{i=1}^{n}\left(I'_{t}(i)-\mu_{i}\right)^{2}\right)
\end{equation}
\smallskip{}

\subsubsection{Visual analysis}

While quantitative network analysis plays a vital role in measuring
specific aspects of interest in a precise and comparable fashion,
network visualization can provide useful overview and exploratory
capabilities, communicating general structure as well as local patterns
of connectivity. The visual analytics paradigm aims at supporting
analytical thinking through interactive visualization, where interaction
is the operative term \cite{keim2008visual}. Through a tight integration
between the user and the data model, users are enabled to explore
and reason about the data. In the case of our dynamic networks, interaction
capabilities for navigating between cross sections and further exploring
network structure provide a setting for qualitative analysis of the
information-rich models.

Force-directed layouting is often used to apply spatialization of
network nodes, that is, to place the nodes in a way that overall approximates
node distances to their corresponding link strengths, thereby seeking
to uncover the structure of the network in terms of more and less
densely connected areas and their relation. Still, force-directed
layouts quickly turn uninformative or ambiguous as the networks become
too dense, including cases of weighted networks with few strong but
many weak connections. Although network visualization with force-directed
layouting often does not scale well to analysis of big networks, it
still can be a useful tool when used properly. In the case of our
bank co-occurrence network it produces decent visualizations for cross
sections of the data set, while stricter filtering of co-occurrences
will produce a more sparse network that is less cluttered. We use
the D3 force algorithm \cite{bostock2011d3} for layouting.

The dynamics of the network can be studied by visualizing cross-sectional
networks in a series, where the positioning is initialized by the
previous step and optimized according to the current linkage, as to
provide continuity that helps in the visual exploration of network
evolution. User interaction plays a vital role not only by allowing
to navigate across time, but also by allowing interaction with the
positioning algorithm, letting the user acquire a more direct understanding
of the structures and details in the data. Force-directed layouting
on more densely linked networks generally finds a locally optimal
positioning out of a large number of comparably good solutions. Interaction
that lets the user drag nodes to reposition them and a force-directed
algorithm that helps to counter-optimize the positioning immediately
afterwards gives rise to a collaborative, exploratory way of working
with and understanding the data.

Nevertheless, the best setting for visual analysis might be one that
combines with quantitative analysis, encoding them visually. For instance,
centrality measures can be encoded by node size to enhance the communication
of structure provided by network visualization, which can use force-directed
layouting or other more regularly structured layouts.

\section{Centrality: Quantitative and qualitative analysis}

This section describes the co-occurrence networks from both a viewpoint
of quantitative measures and exploratory visualization. Starting with
network measures, we describe network properties in general and information
centrality in particular. Then, we turn to visual analysis of the
networks and their constituents.

\subsection{Quantitative analysis}

The volume of bank occurrences is rather stable, apart from a peak
centered around 2008Q1 and some fluctuation from 2012 onward. In 2008
the peak in occurrence volume coincides with a peak in total article
volume, unlike later during the studied time span when occurrence
seems less affected by fluctuating article volume. Interestingly,
the 2008 surge in occurrences barely translates into a rise in co-occurrences
(or strength), i.e., even though banks are more discussed at the time
prior to the outbreak of the crisis, they are not discussed considerably
more in close connection to each other. Overall, total article volume
and bank occurrences have a Pearson correlation of 0.745. Occurrences
and co-occurrences have a correlation of 0.835, which indicates that
there is a notable component to co-occurrence volume which is not
simply explained by occurrence volume.

\begin{figure}
\begin{centering}
\includegraphics[width=0.5\textwidth]{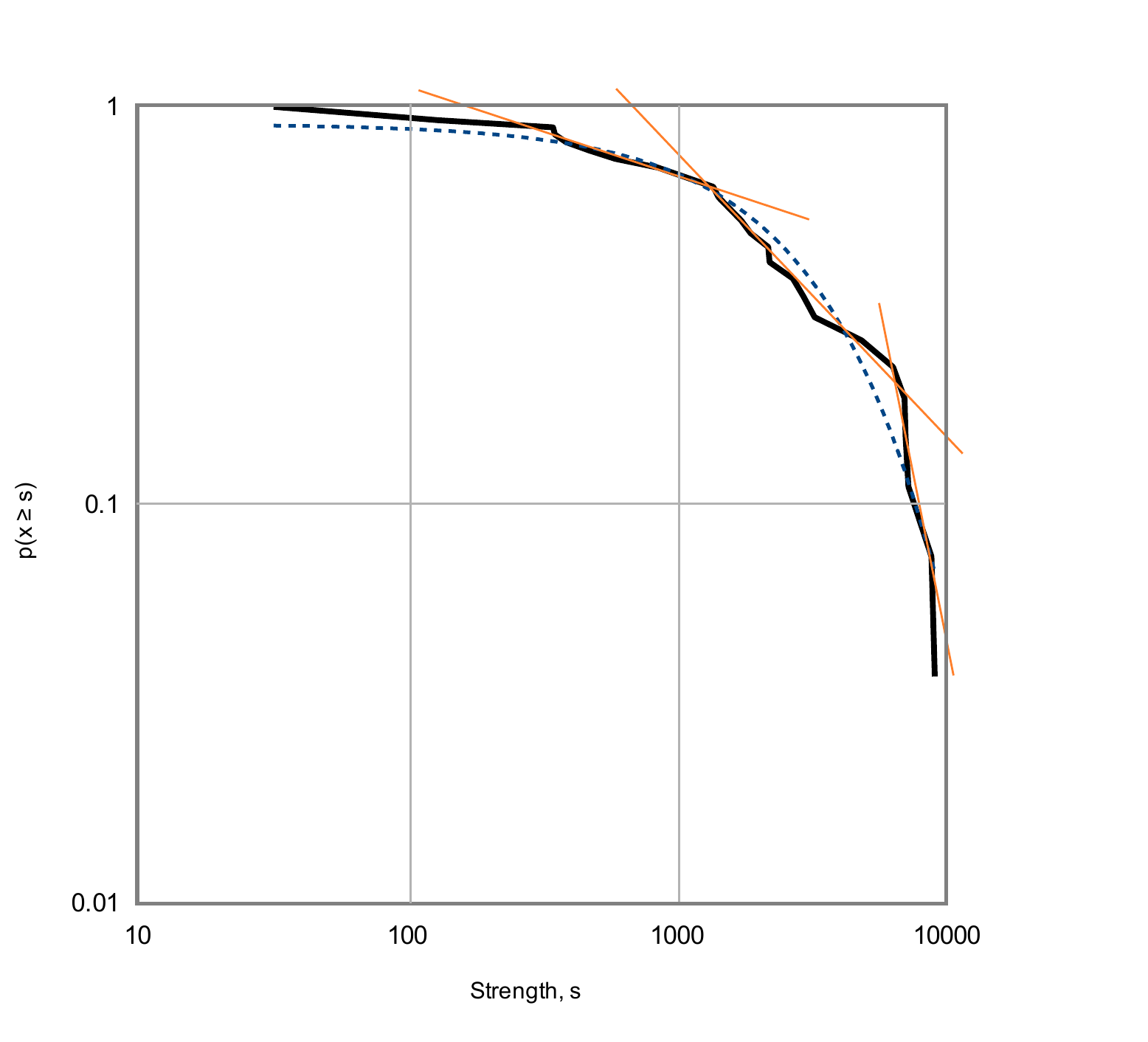}
\par\end{centering}

\protect\caption{Cumulative strength distribution (weighted degree) of bank co-occurrence
network during 2007Q1--2014Q3, showing probability \emph{p} over node
strength \emph{x} against the upper bound on strength \emph{s}. Dashed
line is a fitted exponential function. Solid straight lines indicate
locally fitting power-law functions.\label{fig:Strength-distribution-(weighted}}
\end{figure}

From these aggregated counts, we continue by studying the data as
a network. As discussed in Section 2.3, empirical networks are typically
profiled through measures describing certain global properties. The
average distance, in terms of number of links, between nodes in the
co-occurrence networks are certainly small, at 1.1--1.3, and would
justify calling them 'small-world' networks. However, with weighted
links, a measure of average distance becomes hardly interpretable.
While it is clear that our networks are very tightly connected, the
strength distribution depicts the relative differences in node connectivity.
Many empirical networks exhibit power-law distributed degree or strength,
as a sign of evolution through preferential attachment. Figure \ref{fig:Strength-distribution-(weighted}
shows the cumulative strength distribution of the aggregated network
for the entire period, as well as a closely fitted exponential function
that hints that our network is exceedingly a product of evolution
through uniform attachment. Still, we are able to partially fit power-law
functions to the distribution, as the figure highlights with straight
lines, which could indicate a hybrid model with a weak preferential
attachment component as well. The strength distribution illustrates
the high heterogeneity of connections in the network, i.e., some banks
are much more associated in discussion than others. However, in order
to gain a deeper understanding of a bank's importance to the wider
network, we need to look beyond immediate connections as measured
by degree/strength distribution or degree/strength centrality (proportional
to co-occurrence volume), namely we need to look at information centrality.

\begin{figure}
\begin{centering}
\includegraphics[width=0.95\textwidth]{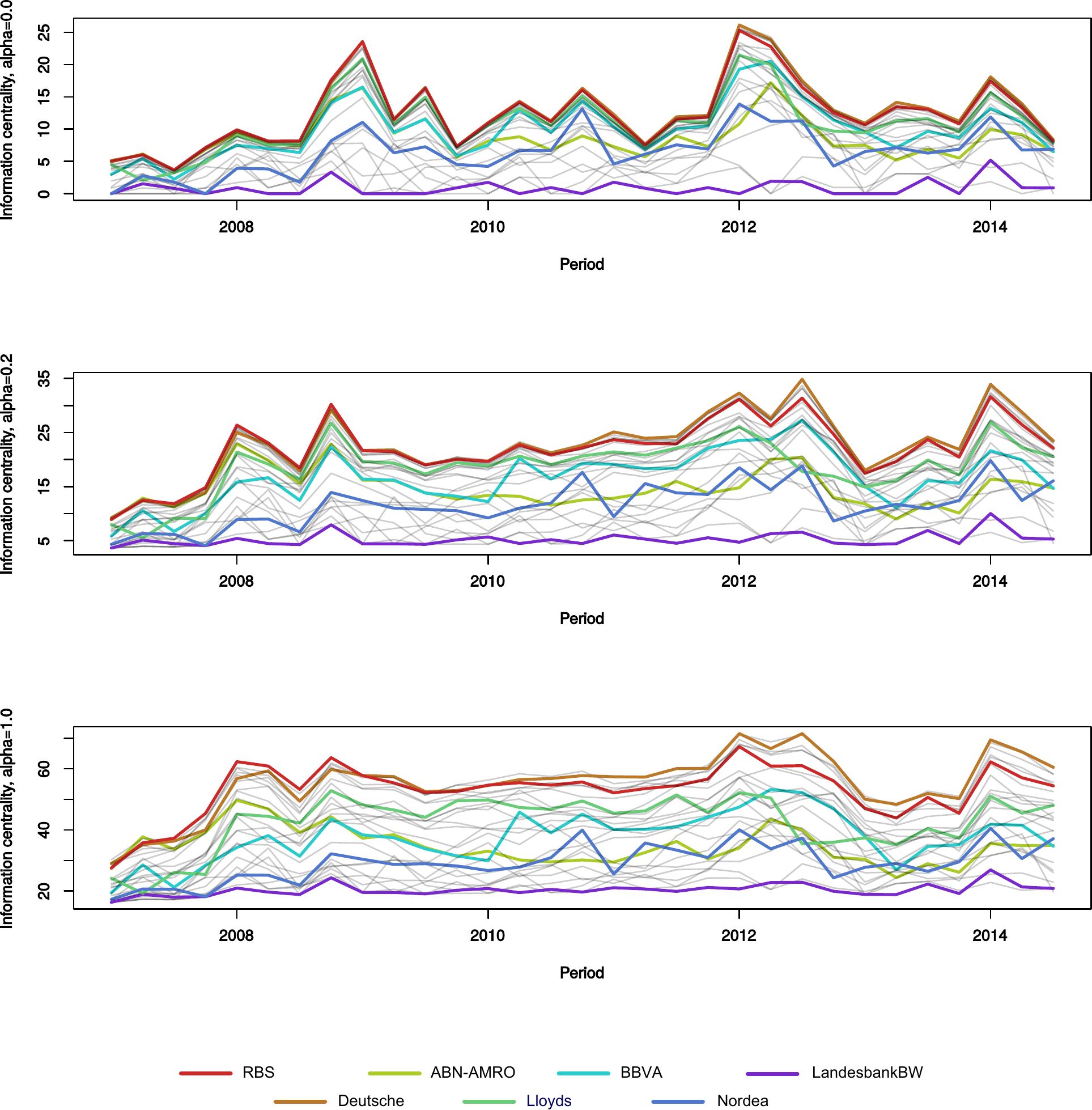}
\par\end{centering}

\bigskip{}

\protect\caption{Information centrality for banks over time. The charts show different
levels of smoothing: none ($\alpha=0.0$), little ($\alpha=0.2$)
and moderate ($\alpha=1.0$). A few example banks are highlighted
(bank labels are described in Table A.1 in the Appendix).\label{fig:Information-centrality-for-time-series}}
\end{figure}

We study information centrality for each node over time, using different
levels of Laplace smoothing (ranging from 0.0 to 5.0). Figure \ref{fig:Information-centrality-for-time-series}
plots the information centrality values, with a number of example
banks highlighted in color and representative $\alpha$ values. Comparing
information centrality with and without smoothing visually, we see
that different peaks are pronounced: some minor peaks (e.g., during
2009Q2-2011Q4) subside while others (e.g., prior to 2008Q3 and crisis
breakout) are substantially amplified even at low levels of smoothing.
Based on its rationale, we interpret that smoothing helps highlight
meaningful patterns in information centrality dynamics and generally
stabilize the series, while reducing artifacts of changing network
size. At higher levels (e.g., $\alpha=1.0$), peaks become relatively
weaker as the distribution of banks evens out on the information centrality
scale, so that fewer banks flock at the very top. We aim to measures
these respective qualities as \emph{V} and \emph{V'} in Equations
2 and 3.

The average variance over time \emph{V} is stationary for very low
values of $\alpha$, with the expected decrease starting at $\alpha=0.2$
(11\% drop from unsmoothed \emph{V}=0.027) and continuing monotonously
with stronger smoothing, directly reflecting its stabilizing nature.
Meanwhile, \emph{V'} signals an increased spread among banks already
at $\alpha=0.01$ (21\% over unsmoothed \emph{V}'=0.025), which reaches
a maximum at $\alpha=1.0$ (94\% increase). We conclude that $\alpha$
levels at or slightly above 0.2 are suitable to achieve moderate smoothing
that communicates global changes of centrality in this network, whereas
1.0 appears to be the optimal choice when focusing on relative differences
in centrality among banks. The regressions in Section 4 use information
centrality with smoothing at $\alpha=1.0$, since relative differences
in centrality are of particular interest. In our experiments, we also
note that\emph{ V} closely follows measures of average covariance
of banks over time, supporting the observation that stronger smoothing
reduces the originally very strong correlation among the most central
banks.

Finally, to test smoothing in relation to sample size, we compare
the variance measures when applied to the above discussed 45\% sample
set to a 20\% sample. As Laplace smoothing is a method to mitigate
effects of limited sample sizes, we expect relatively stronger effects
when applied to a smaller sample. Indeed, $\alpha=0.1$ results in
a 47\% drop in \emph{V} (from 0.03 at $\alpha=0.0$) at 20\% sampling,
while higher alpha only has marginal decreasing impact. Even a small
$\alpha$ has a strong stabilizing effect on the smaller sample, which
in this case contains 1.3M articles. This underlines the fact that
working with text data, typically involving very long-tailed distributions,
often benefits from big data in terms of size to achieve reliable
results, and that smoothing methods are practical for that very reason.
In addition, we tested the robustness of information centrality over
10 random samplings (at 30\%, $\alpha=\{0.0,\,0.1,\,1.0\}$) that
resulted in standard deviations (relative to the mean centrality of
that smoothing level) of 21.2\%, 8.5\% and 3.7\% respectively, which
further highlights the stabilizing effect of smoothing on sparse data.

\begin{figure}
\begin{centering}
\includegraphics[width=0.8\textwidth]{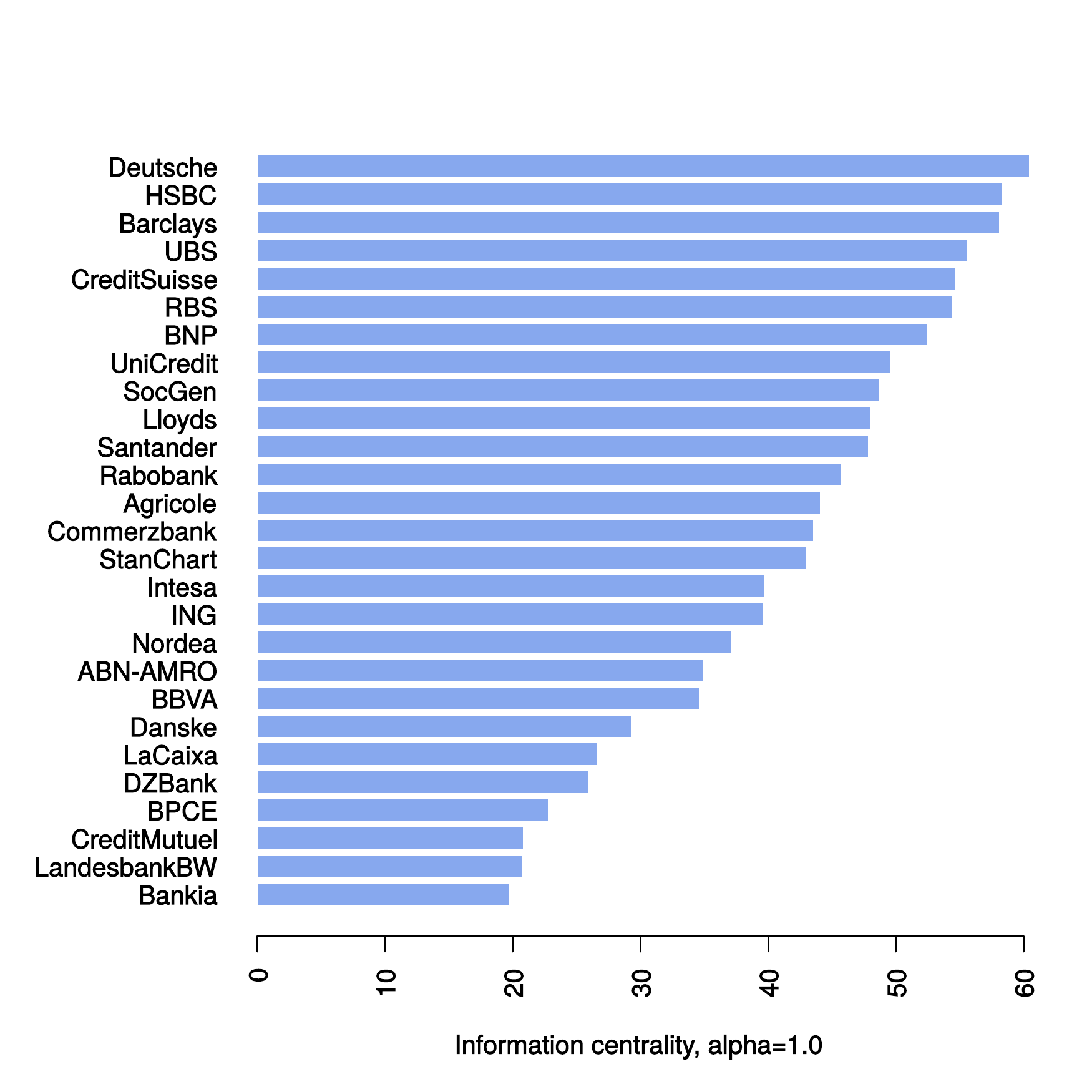}
\par\end{centering}

\protect\caption{Information centrality ranking for all banks in 2014Q3 (bank labels
are described in Table A.1 in the Appendix).\label{fig:Information-centrality-ranking}}
\end{figure}

The trends of individual banks generally follow the movements of the
cross section closely, as increased connectivity in parts of the network
strongly affects the rest, since the co-occurrence network is generally
very tightly connected. Individual centrality relative to the cross
section is generally quite stable. Nevertheless, some changes can
be observed that might reflect real-world events. For instance, ABN
AMRO has relatively high information centrality in 2007 that decreases
afterwards. Royal Bank of Scotland is the most central bank in 2007--2008,
whereas it later on is overtaken by, e.g., Barclays and Deutsche Bank.
To illustrate the information centrality ranking between banks in
more detail, Figure \ref{fig:Information-centrality-ranking} shows
all values as of 2014Q3.

The smoothed information centrality plots exhibit peaks in both 2008Q1
and 2008Q4, as well as during 2012 and 2014Q1. In 2008Q1, for instance,
the peak coincides with the peak in bank occurrence. The fact that
co-occurrence stays relatively flat during the same time indicates
that the change in information centrality is not so much due to generally
strengthened connections, but largely due to change in topology. The
peak in the fourth quarter likewise hints at topological shifts following
the crisis outbreak, but in this period even bank occurrence is normal.
Centralities rise toward 2012, but have subsided substantially in
2013, then coinciding with a similar sharp decrease in bank occurrence.
Overall, the correlation between co-occurrence volume and raw information
centrality averaged over all nodes is 0.651, hinting at a considerable
component other than general co-occurrence volume that we argue is
topological, i.e., involving changes of weight distribution over links
as well as changes in link structure.

\subsection{Visual analysis}

\begin{figure}
\begin{centering}
\includegraphics[height=0.3\textheight]{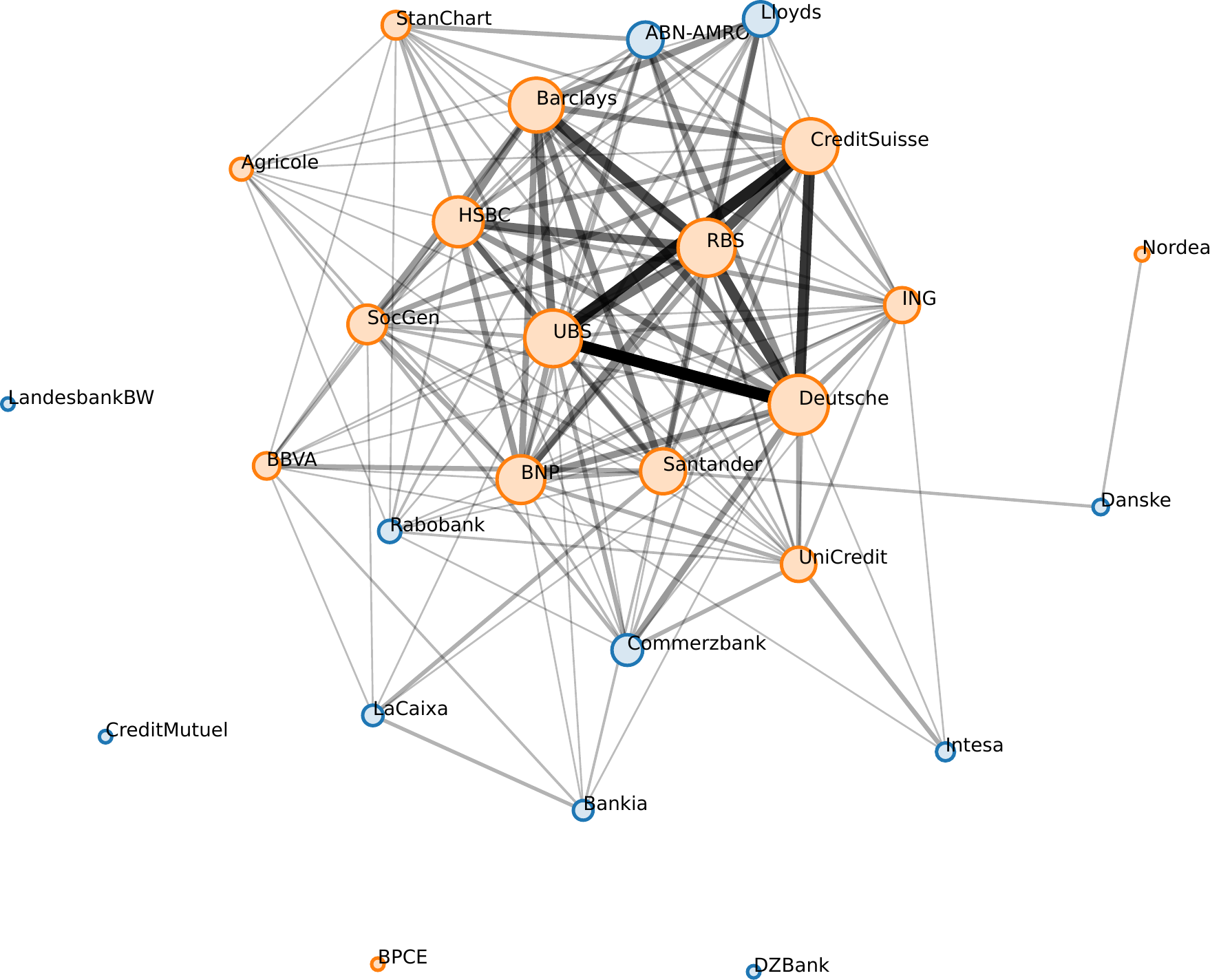}\hfill{}\includegraphics[height=0.3\textheight]{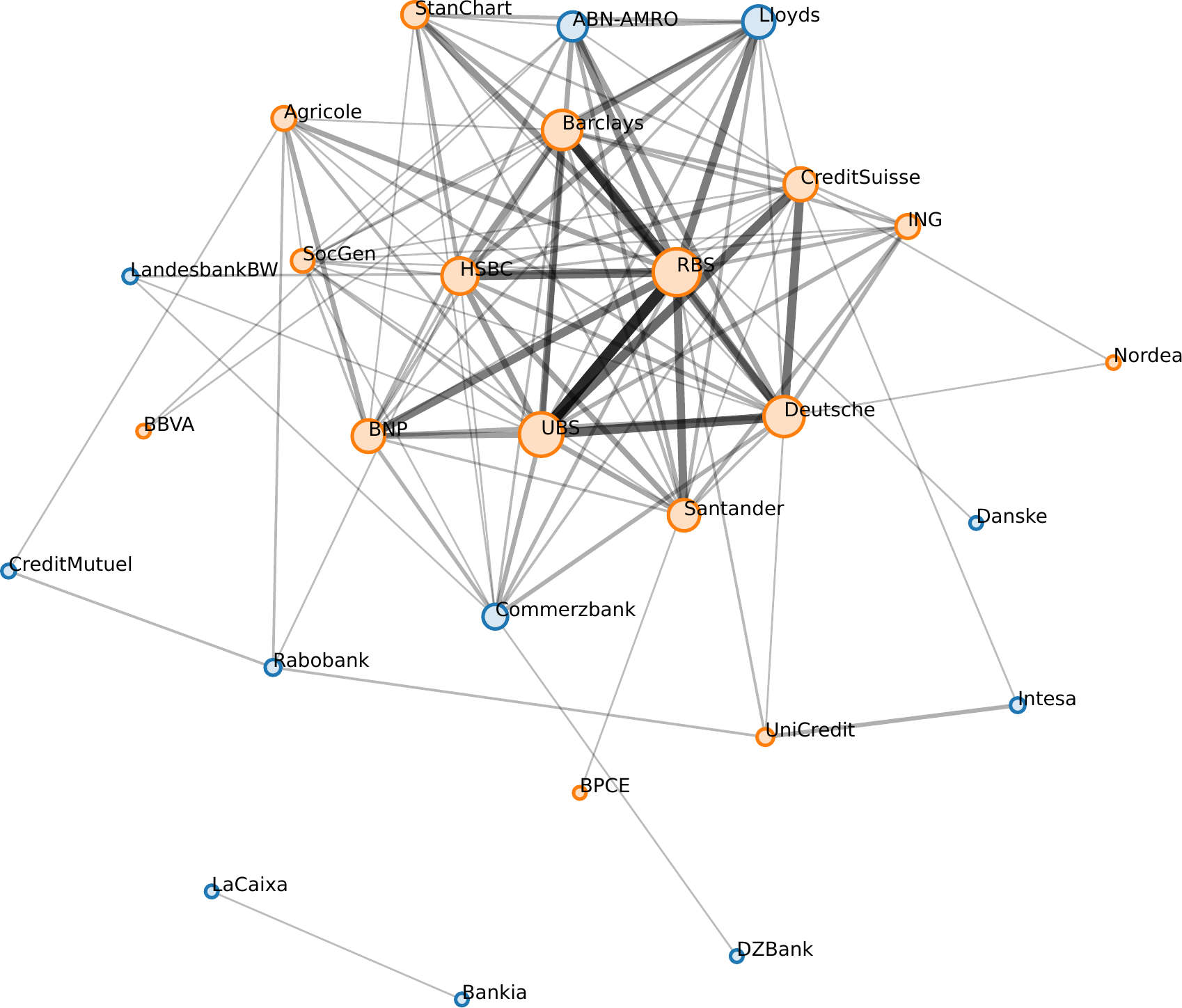}
\par\end{centering}

\medskip{}

\begin{centering}
\hfill{}(a) 2008Q2\hfill{}\hfill{}\hfill{}(b) 2008Q3\hfill{}
\par\end{centering}

\bigskip{}

\begin{centering}
\includegraphics[height=0.3\textheight]{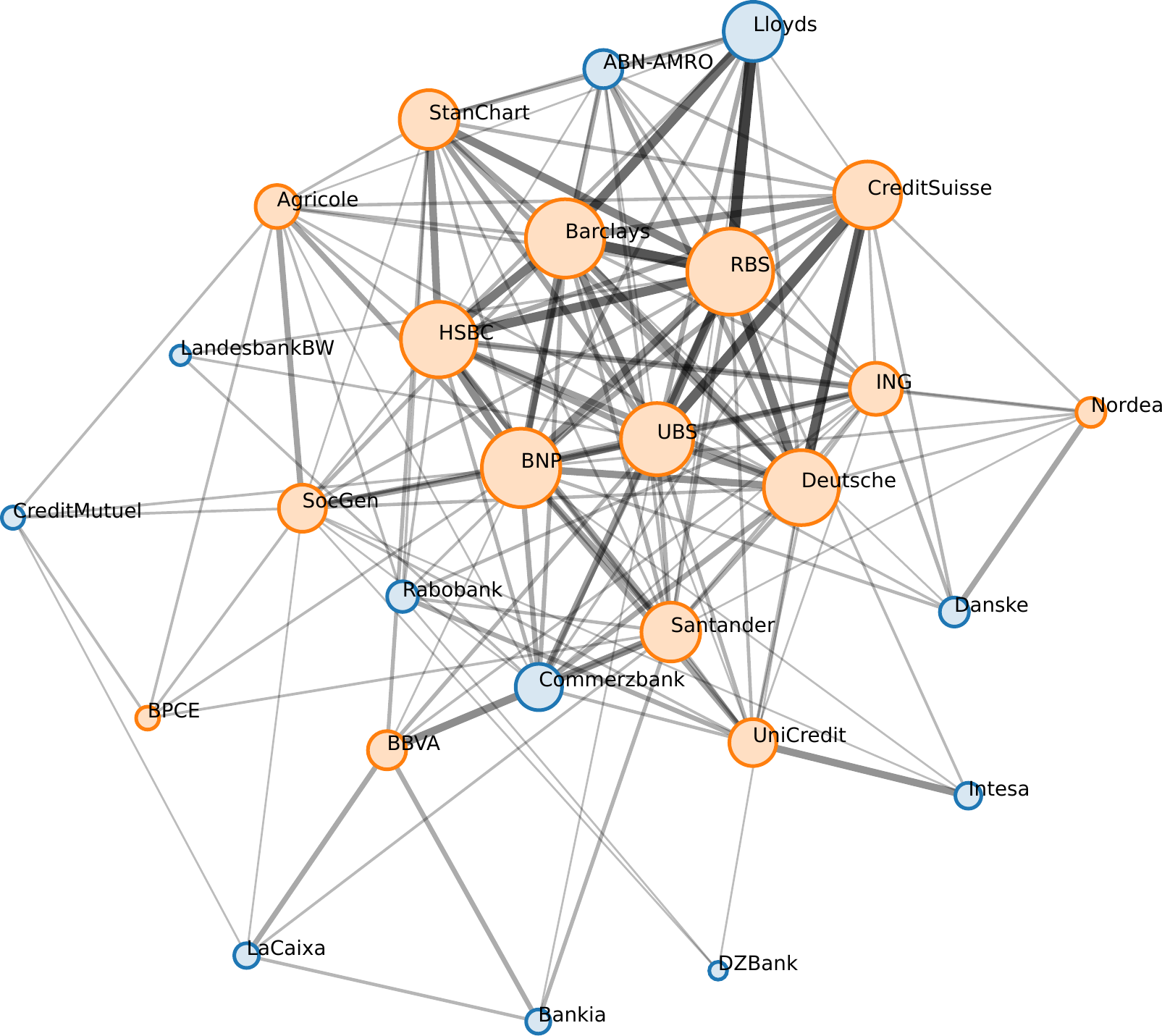}
\par\end{centering}

\medskip{}

\begin{centering}
(c) 2008Q4
\par\end{centering}

\protect\caption{\label{fig:Network-visualization-for}Network visualization for 2008Q2--Q4,
each showing current topology and link strengths (encoded as opacity
and logarithmically scaled line width). Node size is relative to information
centrality ($\alpha=0.2$) and orange color denotes globally systemically
important banks (bank labels are described in Table A.1 in the Appendix).}
\end{figure}
As a complement to the discussion on quantitative analysis of the
co-occurrence networks, we briefly consider the role of visual network
analysis. Our information centrality measurements highlight an interesting
pattern in 2008Q2--Q4 that we inspect further visually. The second
and fourth quarters have relatively high global information centrality,
whereas there is a temporary dip in the third quarter. The networks
in Figure \ref{fig:Network-visualization-for} show visualized snapshots
of each quarter, where the changes in patterns of connectivity can
be studied in more detail. It shows a sparser topology for Q3 than
in both Q2 and Q4, as reflected by the measure. In addition, the visualization
allows for studying local patterns, e.g., how the connection between
the two Scandinavian banks Nordea and Danske Bank (right-side edge
of networks) changes.

In general, the networks have a core consisting of the most central
banks that does not change drastically over time. The periphery experiences
more topological change, but its banks stay mostly in their outside
positions. Nevertheless, it is hardly possible to define a strict
border between core and periphery, neither by visual inspection nor
quantitatively (e.g., by degree or information centrality), rather
the nodes appear on a continuum of centrality (cf. Figure 5). We may
interpret from the force-directed visualization that the network consists
of one major module, with the only exception of occasionally disconnected
components or single nodes (e.g., La Caixa and Bankia in Figure 6b).
The network is overall very densely connected in terms of binary links.

Even though visual inspection can provide valuable insight, in many
cases, it may be hard to reliably and precisely compare changes in
specific aspects, such as centrality of single nodes or centralization
of the whole network, based on the network visualization. This underlines
the importance of backing visual analysis with quantitative measures,
for instance, by encoding node size with information centrality or
presenting plots of measures in parallel, coordinated views. The combination
of both approaches is posed to provide the best possibilities for
understanding the properties of the network, through a mixed process
of exploration and focused inspection. The visual representations
in Figure \ref{fig:Network-visualization-for} represent information
centrality as node size, which in combination with the force-directed
node positioning provides support for visually assessing node centrality
in more general terms.

\section{Determinants of information centrality}

Analysis thus far attempted to convince that information centrality
captures the notion of system-wide importance of a bank in terms of
financial discourse. Yet, little was done to provide a deeper interpretation
of what information centrality signifies. This section explores potential
determinants of information centrality. We explain centrality with
a large number of bank-specific risk drivers, as well as country-specific
macro-financial and banking sector variables, beyond controls for
bank size. Further, we also assess the extent to which information
centrality explains banks' risk to go bad, and compare it to more
standard measures of size.

\subsection{Data}

We complement the textual data, and therefrom derived centrality measures,
with bank-level data from financial statements and banking-sector
and macro-financial indicators at the country level. This gives us
a dataset of 24 risk indicators, spanning 2000Q1 to 2014Q1 for 27
banks, as well as distress events. The definitions of distressed banks
follows Betz et al. \cite{Betzetal2013} and are defined based upon
the following three categories of events: 
\begin{itemize}
\item Direct bank failures include bankruptcies, liquidations and defaults.
\item Government aid events comprise the use of state support on the asset
side, such as capital injections or participation in asset relief
programs (i.e., asset protection or asset guarantees). 
\item Forced mergers capture private sector solutions to bank distress by
conditioning mergers with negative coverage ratios or a parent receiving
state aid after a merger.
\end{itemize}
To measure risk drivers, we make use of CAMELS variables (where the
letters refer to Capital adequacy, Asset quality, Management quality,
Earnings, Liquidity, and Sensitivity to Market Risk). The Uniform
Financial Rating System, informally known as the CAMEL ratings system,
was introduced by the US regulators in 1979. Since 1996, the rating
system was complemented with Sensitivity to Market Risk, to be called
CAMELS. The literature on individual bank failures draws heavily on
the risk drivers put forward by the CAMELS framework. Further, we
complement bank-level data with country-level indicators of risk.
One set of variables describes the banking sector as an aggregate,
whereas another explains macro-financial vulnerabilities in European
countries, such as indicators from the scorecard of the Macroeconomic
Imbalance Procedure. All bank-specific data are retrieved from Bloomberg,
whereas country-level data comes mainly from Eurostat and ECB MFI
Statistics.

\subsection{What explains information centrality?}

The essential question we ask herein is whether more central banks
perform or behave differently. Following Bertay et al. \cite{Bertayetal2013},
who assess whether and to what extent performance, strategy and market
discipline depend on standard bank size measures, we conduct experiments
in order to better understand what signifies information centrality.
In contrast to their study, we control for more standard measures
of bank size, in order to capture particular effects of information
centrality. Using the above described data, we make use of standard,
linear least squares regression models to conduct the following experiments
(cf. Table \ref{tab:Regression-estimates-on}):
\begin{enumerate}
\item Explain information centrality (IC) with bank size variables (Model
1).
\item Explain IC with CAMELS variable groups one-by-one, controlling for
bank size (Models 2--7).
\item Explain IC with all CAMELS variables, controlling for bank size (Model
8).
\item Explain IC with CAMELS and country-specific variables, controlling
for bank size (Model 9).
\end{enumerate}
Our experiments show a number of patterns about drivers of information
centrality. Table \ref{tab:Regression-estimates-on} summarizes all
regression estimates. First, we show that size measures of total assets
and total deposits statistically significantly explain information
centrality. This holds both when included individually and together
in regressions. At a 0.1\% level, we can show that these size variables
relate to centrality, which is in accordance with the nature and aim
of the measure.

Second, we also add variable groups from the CAMELS framework to assess
which risk factors explain information centrality. When testing groups
one-by-one, we find that equity to assets, cost-to-income ratio and
net-short term borrowing are statistically significant at the 5\%
level, and loan loss provisions to total loans, reserves to impaired
assets, interest expenses to liabilities and deposits to funding are
statistically significant at the 1\% level. Large cost-to-income ratios
are expected to reduce individual bank risk, whereas loan loss provisions
are expected to increase risk. Yet, the estimates of the liquidity
variables -- interest expenses to total liabilities and deposits to
funding -- indicate less risk, as more deposits is expected to be
negatively and more interest expenses positively related to bank risk.
The relationships of loan loss reserves and share of trading income
are potentially ambiguous, as higher reserves should correspond to
a higher cover for expected losses but could also proxy for higher
expected losses and trading income might be related to a riskier business
model as a volatile source of earnings but investment securities are
also liquid, allowing to minimize potential fire sale losses.

Third, when including all size and CAMELS variables, we still find
the same variables to be statistically significant, except for all
variables significant at the 5\% level (i.e., equity to assets, cost-to-income
ratio and net-short term borrowing). When assessing the size variables,
assets is consistently a significant predictor, whereas deposits turns
insignificant in Model 6 when also including deposits to funding,
which is likely to be a result of multicollinearity. Further, the
effects of individual risk indicators are unchanged when excluding
all bank size variables, except for slight changes in significance
levels. Fourth, we complement the bank-specific model with country-level
data by also explaining centrality with banking sector and macro-financial
variables. Even though this leads to an improvement of $\mbox{R}^{2}$
by one third, this leaves most of the effects unchanged. Notably,
liquidity indicators and the cost-to-income ratio remain statistically
significant. Out of the country-specific variables, statistically
significant predictors are assets to GDP, non-core liability growth,
loans to deposits, inflation, stock price growth, and sovereign bond
yields.

\begin{sidewaystable}
\protect\caption{Regression estimates on determinants of information centrality\label{tab:Regression-estimates-on}}

\includegraphics[width=1\textwidth]{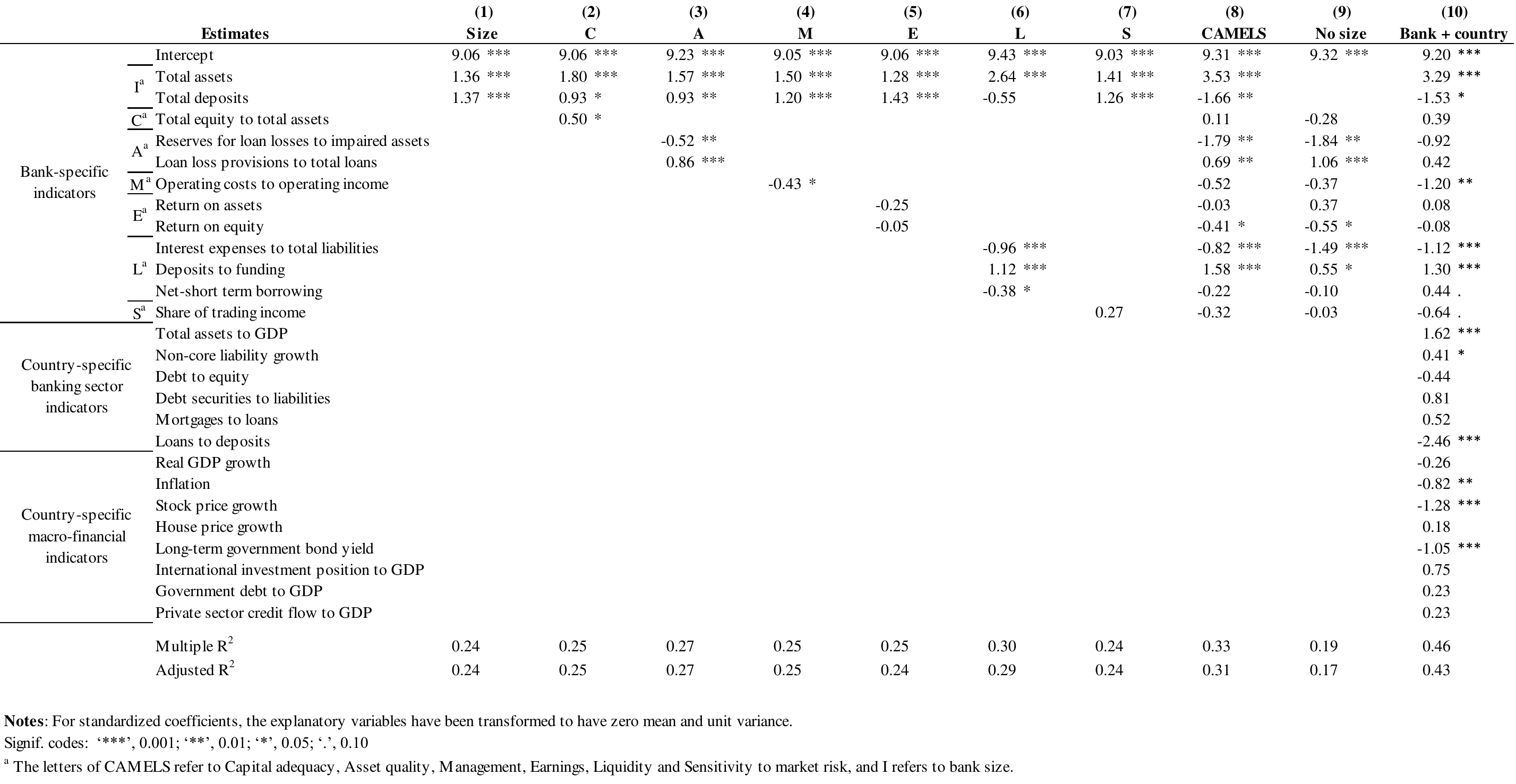}
\end{sidewaystable}

\subsection{Information centrality as a risk driver}

In the above experiments, we showed that information centrality is
partly driven by CAMELS variables, which generally represent different
dimensions of individual bank risk. This does not, however, necessarily
imply that information centrality is a measure of vulnerability. The
next question is whether and to what extent information centrality
signals vulnerable banks, particularly when controlling for CAMELS
variables.

\begin{sidewaystable}
\protect\caption{Early-warning models with information centrality\label{tab:EWM-estimates}}

\includegraphics[width=1\textwidth]{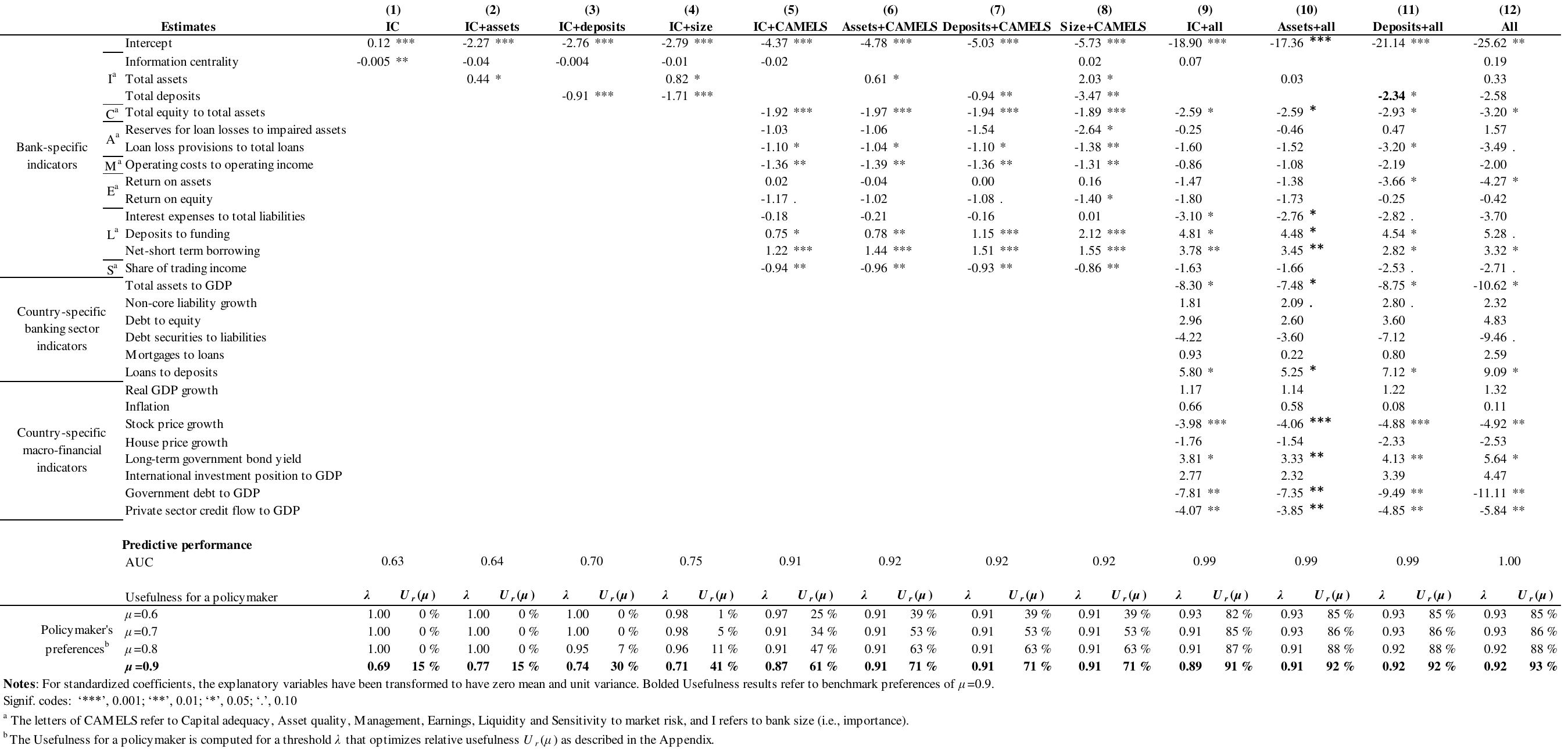}
\end{sidewaystable}

As we have distress events for the banks, and the above used risk
indicators, we can easily test the extent to which information centrality
aids in identifying vulnerable banks. By focusing on vulnerable rather
than distressed banks, we are interested in periods that precede distress
events (e.g., 24 months). In this case, we make use of standard logistic
regression to attain a predicted probability for each bank to be vulnerable.
This probability is turned into a binary point forecast by specifying
a threshold above which we signal vulnerability. This threshold is
chosen to minimize a policymaker's loss function, who has relative
preferences between false alarms and missed crises. Also, we provide
a so-called Usefulness measure that captures the performance of the
model in comparison to not having a model (i.e., best guess of a policymaker).
We assume in the benchmark case the policymaker to be more concerned
about missing a crisis than giving a false alarm, which is particularly
feasible for internal signals. See Appendix B for more details of
the evaluation measures.

To test to what extent information centrality signals vulnerabilities,
and how it relates to bank size variables, we regress pre-distress
events. Hence, as in a standard early-warning setting for banks, we
explain periods 24 months prior to distress with logistic regression.
Starting out from bank importance variables, we can see in Table \ref{tab:EWM-estimates}
(Models 1--4) that while none of the variables yield highly valuable
predictions, assets and deposits provide more Usefulness than information
centrality, particularly deposits. The same holds also for statistical
significance. Even though the bank size variables were above shown
to explain information centrality, we can observe a difference in
their relation to risk. Large banks in terms of assets are found to
be more vulnerable to distress, whereas large banks in terms of deposits
are found to be less so. This is likely to proxy for the business
model or activities of a bank, which might be less risky when the
focus is on depository functions. Moreover, deposits can be seen as
a more stable funding source than interbank market or securities funding.
This points to information centrality being a more general measure
of interconnectedness, rather than one defined by the underlying focus
of the business model. Further, when we add all CAMELS variables to
the three importance measures (Models 5--8), both Usefulness and statistical
significance points to better explanatory power of assets and deposits.
Comparing to models with only bank importance variables, this moves
Usefulness from $U_{r}(\mu=0.9)=41\text{\%}$ at its maximum to $61\text{\%}$
for information centrality and $71\text{\%}$ for assets and deposits.
Likewise, when adding all country-specific variables (Models 9--12),
we can still observe that the explanatory power of assets and deposits
is higher than that for information centrality. At this stage, we
have early-warning models that capture most of the available Usefulness,
by showing a $U_{r}(\mu=0.9)\geq91\text{\%}$. When assessing performance
with the Area Under the Receiver Operating Characteristic Curve (AUC)
(see Appendix B for a description), we can see that the same conclusions
with respect to performance hold.

The implication of the two conducted experiments jointly is that information
centrality is highly correlated with bank size, both when measured
in total assets and deposits, but not a measure of vulnerability.
This indicates that the measure is not biased by business activities
or models, which might be a factor impacting the vulnerability of
a bank. Rather, we are capturing more broadly importance of a bank
in terms of information connectivity in financial discourse. This
property, while due to its broad nature may be a disadvantage, provides
ample means for measuring interconnectedness and centrality from a
wider perspective. It is worth remembering that these text-based networks
are not an ending point, but allow further exploration of the semantics
of observed connections.

\section{Conclusions}

The ongoing global financial crisis has brought interdependencies
among banks into focus in trying to assess interconnected and systemic
risk.This paper has demonstrated the use of computational analysis
of financial discussion, as a source for information on bank interrelations.
Conventional approaches make use of direct linkages to the extent
available and market-based measures as an indirect estimate of interdependence,
which both have their limitations, such as non-publicly disclosed
information, strong ties to specific business models and deficiencies
in the forward-lookingness of co-movements in markets. The approach
we put forward may serve as a complement to more established ways
of quantifying connectedness and dependence among banks. We have presented
a text-to-network process, which has its basis in co-occurrences of
bank names and can be analyzed quantitatively and visualized. To support
quantification of bank importance, we proposed an information centrality
measure to rank and assess trends of bank centrality in discussion.
Rather than a common shortest-path-based centrality measure, information
centrality captures effects that might propagate aimlessly by accounting
for parallel paths. Moreover, we proposed a method to stabilize the
centrality measure by applying Laplace smoothing to link weights before
calculating information centrality. To support a qualitative assessment
of the bank networks, we put forward a visual, interactive interface
for better illustrating network structures. This concerned not only
an interface to network models, but also an interactive plot to better
communicate quantitative network measures. Our text-based approach
was illustrated on European Large and Complex Banking Groups (LCBGs)
during the ongoing financial crisis by quantifying bank interrelations
from discussion in 3.0M news articles, spanning the years 2007 to
2014 (Q3).

To better understand the interpretation of and what drives information
centrality, we have explored determinants of the centrality measure.
We investigated bank-specific and country-specific risk drivers, as
well as control for variables measuring bank size, and also assess
the extent to which bank risk is explained by information centrality
in relation to more standard measures of size. We have shown that
centrality is not a direct measure of vulnerability, despite the fact
that it is closely linked to size variables. The conclusions to be
drawn from this are that the centrality measure is not biased by the
nature of business activities or models, which may impact the vulnerability
of banks (e.g., asset size or interbank-lending centrality). Instead,
the measure of information centrality is described to capture the
importance of a bank in a wider perspective, in terms of information
connectivity in financial discourse.

Considering the limitations of the current network and that the underlying
data occasionally lead to somewhat hazy patterns difficult to interpret
and draw clear conclusions from, we suggest a number of ways these
issues could be addressed in future research. One advantage of using
text data is the potentially rich semantic information it holds, which
can be used to better explain or narrow the relations extracted, thereby
facilitating interpretation of the network and the measures applied
on top. A disadvantage of applying such filtering might be that it
vastly increases the data size requirements, quickly reducing a big
but sparse data set into a rather scarce one. Similarly, in this paper
we have illustrated how, using a data set of a few million articles,
accuracy can still be improved by even more data. 

Although textual data provides the basis for studying interrelationships
and other potentially interesting details on banks more specifically,
its interpretation by computational methods is often challenging.
In order to apply filtering by theme to co-occurrence links between
banks, we recommend more sophisticated semantic analysis to increase
recall. For instance, distributional semantic methods \cite{turney2010frequency}\cite{mikolov2013efficient}
could be used to extend a set of seed keywords, or probabilistic topic
modeling \cite{blei2012probabilistic} could be applied to the corpus
to identify topics of interest and the related subset of articles.
Furthermore, combining sentiment analysis with our bank relation extraction
could constitute another interesting way to distinguish the nature
of mapped relations. Sentiment analysis has been applied to classify
company-related information from financial news in regards to the
effect on their stock price (e.g., \cite{malo2013learning}), an approach
that could hold considerable potential in the area of systemic risk
analysis as well.

\pagebreak{}

\section*{References}

\bibliographystyle{plain}
\bibliography{references}

\begin{thebibliography}{10}

\bibitem{acharya2012measuring}
Viral Acharya, Lasse Pedersen, Thomas Philippon, and Matthew Richardson.
\newblock Measuring systemic risk.
\newblock 2012.

\bibitem{barabasi1999emergence}
Albert-L{\'a}szl{\'o} Barab{\'a}si and R{\'e}ka Albert.
\newblock Emergence of scaling in random networks.
\newblock {\em Science}, 286(5439):509--512, 1999.

\bibitem{Barrat16032004}
A.~Barrat, M.~Barthelemy, R.~Pastor-Satorras, and A.~Vespignani.
\newblock The architecture of complex weighted networks.
\newblock {\em Proceedings of the National Academy of Sciences of the United
  States of America}, 101(11):3747--3752, 2004.

\bibitem{Battistonetal2012}
S.~Battiston, M.~Puliga, R.~Kaushik, P.~Tasca, and G.~Caldarelli.
\newblock Debtrank: Too central to fail? financial networks, the fed and
  systemic risk.
\newblock {\em Scientific Reports}, 2:541, 2012.

\bibitem{Bertayetal2013}
A.C. Bertay, A.~Demirg{\"u}{\c{c}}-Kunt, and H.~Huizinga.
\newblock Do we need big banks? evidence on performance, strategy and market
  discipline.
\newblock {\em Journal of Financial Intermediation}, 22(4):532--558, 2013.

\bibitem{Betzetal2013}
F.~Betz, S.~Oprica, T.~Peltonen, and P.~Sarlin.
\newblock Predicting distress in {E}uropean banks.
\newblock {\em Journal of Banking \& Finance}, 45:225--241, 2014.

\bibitem{blei2012probabilistic}
David~M Blei.
\newblock Probabilistic topic models.
\newblock {\em Communications of the ACM}, 55(4):77--84, 2012.

\bibitem{borgatti2005centrality}
Stephen~P Borgatti.
\newblock Centrality and network flow.
\newblock {\em Social networks}, 27(1):55--71, 2005.

\bibitem{borio2009towards}
Claudio~EV Borio and Mathias Drehmann.
\newblock {\em Towards an operational framework for financial stability:
  "fuzzy" measurement and its consequences}.
\newblock Number 284. Bank for International Settlements, Monetary and Economic
  Department, 2009.

\bibitem{bostock2011d3}
Michael Bostock, Vadim Ogievetsky, and Jeffrey Heer.
\newblock D3: Data-driven documents.
\newblock {\em IEEE Trans. Visualization \& Comp. Graphics (Proc. InfoVis)},
  2011.

\bibitem{brandes2005centrality}
Ulrik Brandes and Daniel Fleischer.
\newblock Centrality measures based on current flow.
\newblock {\em STACS 2005}, pages 533--544, 2005.

\bibitem{bunescu2005shortest}
Razvan~C Bunescu and Raymond~J Mooney.
\newblock A shortest path dependency kernel for relation extraction.
\newblock In {\em Proceedings of the conference on Human Language Technology
  and Empirical Methods in Natural Language Processing}, pages 724--731.
  Association for Computational Linguistics, 2005.

\bibitem{cerutti2012systemic}
Eugenio Cerutti, Stijn Claessens, and Patrick McGuire.
\newblock Systemic risk in global banking: what can available data tell us and
  what more data are needed?
\newblock 2012.

\bibitem{chen1999empirical}
Stanley~F Chen and Joshua Goodman.
\newblock An empirical study of smoothing techniques for language modeling.
\newblock {\em Computer Speech \& Language}, 13(4):359--393, 1999.

\bibitem{dell2008real}
Giovanni Dell'Ariccia, Enrica Detragiache, and Raghuram Rajan.
\newblock The real effect of banking crises.
\newblock {\em Journal of Financial Intermediation}, 17(1):89--112, 2008.

\bibitem{dhar2013data}
Vasant Dhar.
\newblock Data science and prediction.
\newblock {\em Communications of the ACM}, 56(12):64--73, 2013.

\bibitem{dijkstra1959note}
Edsger~W Dijkstra.
\newblock A note on two problems in connexion with graphs.
\newblock {\em Numerische mathematik}, 1(1):269--271, 1959.

\bibitem{ECB13fsr}
European Central Bank.
\newblock {\em Financial Stability Review}, November 2013.

\bibitem{FSB2013update}
Financial Stability Board.
\newblock {\em 2013 update of group of global systemically important banks
  (G-SIBs)}, November 11 2013.

\bibitem{hautsch2013financial}
Nikolaus Hautsch, Julia Schaumburg, and Melanie Schienle.
\newblock Financial network systemic risk contributions.
\newblock Technical report, CFS Working Paper, 2013.

\bibitem{jackson2007meeting}
Matthew~O Jackson and Brian~W Rogers.
\newblock Meeting strangers and friends of friends: How random are social
  networks?
\newblock {\em The American economic review}, pages 890--915, 2007.

\bibitem{keim2008visual}
Daniel~A Keim, Florian Mansmann, J{\"o}rn Schneidewind, Jim Thomas, and Hartmut
  Ziegler.
\newblock {\em Visual Data Mining}, chapter Visual analytics: Scope and
  challenges, page~76.
\newblock Springer, 2008.

\bibitem{laeven2010resolution}
Luc Laeven and Fabian Valencia.
\newblock {\em Resolution of banking crises: The good, the bad, and the ugly}.
\newblock International Monetary Fund, 2010.

\bibitem{malkiel2003efficient}
Burton~G Malkiel.
\newblock The efficient market hypothesis and its critics.
\newblock {\em Journal of economic perspectives}, pages 59--82, 2003.

\bibitem{malo2013learning}
Pekka Malo, Ankur Sinha, Pyry Takala, Oskar Ahlgren, and Iivari Lappalainen.
\newblock Learning the roles of directional expressions and domain concepts in
  financial news analysis.
\newblock In {\em Data Mining Workshops (ICDMW), 2013 IEEE 13th International
  Conference on}, pages 945--954. IEEE, 2013.

\bibitem{mikolov2013efficient}
Tomas Mikolov, Kai Chen, Greg Corrado, and Jeffrey Dean.
\newblock Efficient estimation of word representations in vector space.
\newblock 2013.

\bibitem{newman2001scientific}
Mark~EJ Newman.
\newblock Scientific collaboration networks. ii. shortest paths, weighted
  networks, and centrality.
\newblock {\em Physical review E}, 64(1):016132, 2001.

\bibitem{opsahl2010node}
Tore Opsahl, Filip Agneessens, and John Skvoretz.
\newblock Node centrality in weighted networks: Generalizing degree and
  shortest paths.
\newblock {\em Social Networks}, 32(3):245--251, 2010.

\bibitem{ozgur2008co}
Arzucan {\"O}zg{\"u}r, Burak Cetin, and Haluk Bingol.
\newblock Co-occurrence network of reuters news.
\newblock {\em International Journal of Modern Physics C}, 19(05):689--702,
  2008.

\bibitem{RonnqvistSarlin2014}
Samuel R\"onnqvist and Peter Sarlin.
\newblock From text to bank interrelation maps.
\newblock In {\em IEEE Conference on Computational Intelligence for Financial
  Engineering \& Economics (CIFEr)}, 2014.

\bibitem{Sarlin2013b}
P.~Sarlin.
\newblock On policymakers' loss functions and the evaluation of early warning
  systems.
\newblock {\em Economics Letters}, 119(1):1--7, 2013.

\bibitem{Sarlin2013SWIFT}
P~Sarlin.
\newblock Macroprudential oversight, risk communication and visualization.
\newblock {LSE SP Working Paper No. 4.}, 2014.

\bibitem{Soramakietal2007}
K.~Soram\"aki, M.L. Bech, J.~Arnold, R.J. Glass, and W.E. Beyeler.
\newblock The topology of interbank payment flows.
\newblock {\em Physica A: Statistical Mechanics and its Applications},
  379(1):317--333, 2007.

\bibitem{Stephenson19891}
Karen Stephenson and Marvin Zelen.
\newblock Rethinking centrality: Methods and examples.
\newblock {\em Social Networks}, 11(1):1 -- 37, 1989.

\bibitem{turney2010frequency}
Peter~D Turney and Patrick Pantel.
\newblock From frequency to meaning: Vector space models of semantics.
\newblock {\em Journal of artificial intelligence research}, 37(1):141--188,
  2010.

\bibitem{watts1998collective}
Duncan~J Watts and Steven~H Strogatz.
\newblock Collective dynamics of 'small-world' networks.
\newblock {\em Nature}, 393(6684):440--442, 1998.

\bibitem{wren2004knowledge}
Jonathan~D Wren, Raffi Bekeredjian, Jelena~A Stewart, Ralph~V Shohet, and
  Harold~R Garner.
\newblock Knowledge discovery by automated identification and ranking of
  implicit relationships.
\newblock {\em Bioinformatics}, 20(3):389--398, 2004.

\end{thebibliography}

\section*{\textmd{\small{}\newpage{}}}

\section*{Appendix A: Data}

\setcounter{table}{0}

\renewcommand{\thetable}{A.\arabic{table}}

\begin{table}[H]
\protect\caption{A list of banks and their labels.}

~

\noindent \centering{}%
\begin{tabular}{ccccc}
\multicolumn{2}{c}{European LCBG and G-SIB} &  & \multicolumn{2}{c}{European LCBG}\tabularnewline
Label & Name &  & Label & Name\tabularnewline
\hline 
\hline 
{\small{}Agricole} & {\small{}Credit Agricole Groupe} &  & {\small{}ABN-AMRO} & {\small{}ABN AMRO Bank NV}\tabularnewline
{\small{}BBVA } & {\small{}Banco Bilbao Vizcaya Argenta} &  & {\small{}Bankia} & {\small{}Bankia SA}\tabularnewline
{\small{}BPCE} & {\small{}Groupe BPCE} &  & {\small{}Commerzbank} & {\small{}Commerzbank AG}\tabularnewline
{\small{}BNP} & {\small{}BNP Paribas} &  & {\small{}CreditMutuel} & {\small{}Credit Mutuel Group}\tabularnewline
{\small{}Barclays} & {\small{}Barclays PLC} &  & {\small{}DZBank} & {\small{}DZ Bank AG}\tabularnewline
{\small{}CreditSuisse} & {\small{}Credit Suisse Group AG} &  & {\small{}Danske} & {\small{}Danske Bank A/S}\tabularnewline
{\small{}Deutsche} & {\small{}Deutsche Bank AG} &  & {\small{}Intesa} & {\small{}Intesa Sanpaolo}\tabularnewline
{\small{}HSBC} & {\small{}HSBC Holdings PLC} &  & {\small{}LaCaixa} & {\small{}La Caixa}\tabularnewline
{\small{}ING} & {\small{}ING Bank NV} &  & {\small{}LandesbankBW} & {\small{}Landesbank Baden-Württemberg}\tabularnewline
{\small{}Nordea} & {\small{}Nordea Bank AB} &  & {\small{}Lloyds} & {\small{}Lloyds Banking Group PLC}\tabularnewline
{\small{}RBS} & {\small{}Royal Bank of Scotland} &  & {\small{}Rabobank} & {\small{}Rabobank Group}\tabularnewline
{\small{}Santander} & {\small{}Banco Santander SA} &  &  & \tabularnewline
{\small{}SocGen} & {\small{}Group Societe Generale SA} &  &  & \tabularnewline
{\small{}StanChart} & {\small{}Standard Chartered PLC} &  &  & \tabularnewline
{\small{}UBS} & {\small{}UBS AG} &  &  & \tabularnewline
\end{tabular}
\end{table}

\section*{Appendix B: Usefulness of early-warning models}

\setcounter{table}{0}

\renewcommand{\thetable}{B.\arabic{table}}

Early-warning models require evaluation criteria that account for
the nature of low-probability, high-impact events. Following Sarlin
\cite{Sarlin2013b}, the signal evaluation framework focuses on a
policymaker with relative preferences between type I and II errors,
and the usefulness that she derives by using a model, in relation
to not using it. To mimic an ideal leading indicator, we build a binary
state variable $C_{j}(h)\in\left\{ 0,1\right\} $ for observation
$j$ (where $j=1,2,\ldots,N$) given a specified forecast horizon
$h$. Let $C_{j}(h)$ be a binary indicator that is one during pre-crisis
periods and zero otherwise. For detecting events $C_{j}$ using information
from indicators, we estimate the probability of a crisis occurrence
$p_{j}\in\left[0,1\right]$, for which we use herein logistic regression.
The probability $p_{j}$ is turned into a binary prediction $P_{j}$,
which takes the value one if $p_{j}$ exceeds a specified threshold
$\lambda\in\left[0,1\right]$ and zero otherwise. The correspondence
between the prediction $P_{j}$ and the ideal leading indicator $C_{j}$
can then be summarized into a so-called contingency matrix. 

\begin{table}[h]
\protect\caption{A contingency matrix.}

~

\noindent \centering{}%
\begin{tabular}{|c|c|c|c|}
\cline{3-4} 
\multicolumn{1}{c}{} &  & \multicolumn{2}{c|}{{\small{}Actual class $C_{j}$}}\tabularnewline
\cline{3-4} 
\multicolumn{1}{c}{} &  & {\small{}Crisis} & {\small{}No crisis}\tabularnewline
\hline 
\multirow{4}{*}{{\small{}Predicted class $P_{j}$}} & \multirow{2}{*}{{\small{}Signal}} & {\small{}Correct call} & {\small{}False alarm}\tabularnewline
 &  & \emph{\small{}True positive (TP)} & \emph{\small{}False positive (FP)}\tabularnewline
\cline{2-4} 
 & \multirow{2}{*}{{\small{}No signal}} & {\small{}Missed crisis} & {\small{}Correct silence}\tabularnewline
 &  & \emph{\small{}False negative (FN)} & \emph{\small{}True negative (TN)}\tabularnewline
\hline 
\end{tabular}
\end{table}

The frequencies of prediction-realization combinations in the contingency
matrix are used for computing a wide range of quantitative measures
of classification performance. Beyond measures of overall accuracy,
a policymaker can be thought to be primarily concerned with two types
of errors: issuing a false alarm and missing a crisis. The evaluation
framework described below is based upon that in Sarlin \cite{Sarlin2013b}
for turning policymakers' preferences into a loss function, where
the policymaker has relative preferences between type I and II errors.
While type I errors represent the share of missed crises to the frequency
of crises $T_{1}\in\left[0,1\right]=$FN/(TP+FN), type II errors represent
the share of issued false alarms to the frequency of tranquil periods
$T_{2}\in\left[0,1\right]=$FP/(FP+TN). Given probabilities $p_{j}$
of a model, the policymaker then optimizes the threshold $\lambda$
such that her loss is minimized. The loss of a policymaker includes
$T_{1}$ and $T_{2}$, weighted by relative preferences between missing
crises ($\mu$) and issuing false alarms ($1-\mu$). By accounting
for unconditional probabilities of crises $P_{1}=P(C=1)$ and tranquil
periods $P_{2}=P(C=0)=1-P_{1}$, the loss function can be written
as follows:

\begin{equation}
L(\mu)=\mu T_{1}P_{1}+(1-\mu)T_{2}P_{2}
\end{equation}

\hspace{-0.52cm}where $\mu\in\left[0,1\right]$ represents the relative
preferences of missing crises and $1-\mu$ of giving false alarms,
$T_{1}$ the type I errors, and $T_{2}$ the type II errors. $P_{1}$
refers to the size of the crisis class and $P_{2}$ to the size of
the tranquil class. Further, the Usefulness of a model can be defined
in a more intuitive manner. First, the absolute Usefulness ($U_{a}$)
is given by: 

\begin{equation}
U_{a}(\mu)=\mbox{min}(\mu P_{1},\left(1-\mu\right)P_{2})-L(\mu),
\end{equation}

\hspace{-0.52cm}which computes the superiority of a model in relation
to not using any model. As the unconditional probabilities are commonly
unbalanced and the policymaker may be more concerned about the rare
class, a policymaker could achieve a loss of $\mbox{min}(\mu P_{1},\left(1-\mu\right)P_{2})$
by either always or never signaling a crisis. This predicament highlights
the challenge in building a useful early-warning model: with an imperfect
model, it would otherwise easily pay off for the policymaker to always
signal the high-frequency class. 

Second, we can compute the relative Usefulness $U_{r}$ as follows: 

\begin{equation}
U_{r}(\mu)=\frac{U_{a}(\mu)}{\mbox{min}(\mu P_{1},\left(1-\mu\right)P_{2})},
\end{equation}

\hspace{-0.52cm}where $U_{a}$ of the model is compared with the
maximum possible usefulness of the model. That is, the loss of disregarding
the model is the maximum available Usefulness. Hence, $U_{r}$ reports
$U_{a}$ as a share of the Usefulness that a policymaker would gain
with a perfectly-performing model, which supports interpretation of
the measure. 
\end{document}